\documentclass[aps,prl,preprint,groupedaddress]{revtex4-1}

\usepackage{float}
\usepackage{hyperref}
\usepackage{epsfig}
\usepackage{graphicx}
\usepackage{subfigure}
\usepackage{latexsym}
\usepackage{color}
\usepackage{fullpage}
\usepackage{epstopdf}
\usepackage{dcolumn}
\usepackage{bm}
\usepackage{ulem}
\usepackage{units}
\usepackage{amsmath}
\usepackage{lineno}
\usepackage{verbatim}

\begin{document}

\title{Artificial vertex systems by design}

\author{V. Sch\'anilec$^{1,2}$, Y. Perrin$^{1,2}$, S. Le Denmat$^{1,2}$, B. Canals$^{1,2}$ and N. Rougemaille$^{1,2}$}

\affiliation{$^1$ CNRS, Inst NEEL, F-38000 Grenoble, France\\$^2$ Univ. Grenoble Alpes, Inst NEEL, F-38000 Grenoble, France}

\date{\today}

\begin{abstract}

\textbf{Arrays of interacting magnetic nanostructures were introduced recently as a powerful approach to investigate experimentally the exotic many-body physics of frustrated spin models. 
Following a similar strategy based on lithographically-patterned magnetic lattices, we provide a first attempt to fabricate a lab-on-chip platform to explore the physics of vertex models.
The central idea of this work is to replace the spin degree of freedom of artificial frustrated spin systems by a local micromagnetic knob, which can be finely tuned by a proper design of the vertex geometry.
This concept is demonstrated both numerically and experimentally on the celebrated six vertex model, and we show how all variants of this model can be apprehended through the engineering of magnetic square lattices.
Our results open new avenues to design arrays of magnetic nanostructures not only to study a wide range of frustrated spin models, but to explore vertex models as well.}

\end{abstract}

\maketitle

Artificial systems are often used in physics to induce, explore and manipulate intriguing properties of matter, which do not exist in nature or which can be challenging to investigate otherwise \cite{Atomes_froids, Schiffer_review}. 
In condensed matter magnetism, artificial arrays of interacting nanomagnets were introduced \cite{Tanaka2006, Wang2006} as a possible way to fabricate experimentally various types of frustrated spin models. 
Following earlier attempts \cite{Pannetier1993, Pannetier2002, Davidovic1996, Davidovic1997, Hilgenkamp2005}, this idea to use lithographically-patterned architectures, be they magnetic or not, triggered a wealth of studies on frustrated spin systems \cite{Libal2006, Han2008, Qi2008, Morgan2011, Farhan2013, Porro2013, Cumings2014, Kapaklis2014, Tierno2016, Perrin2016, Nguyen2017}. 
Many predictions and observations from condensed matter have been revisited, tested and even extended by using artificial spin systems. 
For instance, these magnetic systems allow to visualize unconventional magnetic phases and exotic collective phenomena directly in real space, in an almost routinely fashion. 
In kagome dipolar spin ice systems, the observation of a magnetic phase in which order and disorder coexist at thermodynamic equilibrium is one example of the intriguing physics that can be probed \cite{Rougemaille2011, Zhang2013, Chioar2014-1, Montaigne2014, Drisko2015, Anghinolfi2015, Canals2016}. 
Because almost any type of two-dimensional geometry can be designed, whether or not this geometry exist in nature \cite{Bhat2013, Phatak2016, Gilbert2014, Gilbert2015}, artificial spin systems offer a wide range of possibilities to investigate exotic magnetic states of matter and complex magnetic ordering \cite{Zhang2012, Chioar2014-2, Chioar2016}, especially in magnetically frustrated spin systems.

In this work, we propose a strategy to extend the use of lithographically-patterned magnetic systems for investigating vertex models with the very same lab-on-chip approach developed for artificial spin systems.
These artificial vertex systems are built from arrays of nanomagnets which are physically connected at the nodes of the lattice such that micromagnetism now plays a key role.
Under such conditions, a non-uniform magnetization distribution appears at the vertices \cite{Zeissler2013, Rougemaille2013}, where the nanomagnets meet.
This magnetization distribution shares common features with magnetic domain walls in magnetic nanostrips, and the energetics of these domain walls results from the competition between the magnetostatic and microscopic exchange interactions.
By adjusting the vertex geometry, the energy of the different possible domain walls can be tuned continuously using micromagnetism as a knob, and the energy hierarchy of the different vertex types can be modified.
More specifically, we focus our work on the six vertex model and demonstrate that the three predicted phases can be reached experimentally, including the macroscopically degenerated manifold of the disordered phase obtained when all six vertices have the same energy.
This result is important as it opens new avenues to investigate in real space a spin liquid state in a vertex system, i.e., a system in which magnetic frustration is not driven by interactions nor by topology, but implemented more deeply, through a local constraint. 
In other words, instead of finding technological strategies to engineer a frustrated spin system into its highly degenerate low-energy manifold, we provide a simple and elegant experimental solution to adjust the vertex energy at will, without the need of finely tune two-body frustrated interactions.

\subsection{The six vertex model}

\begin{figure}[H]
\center
\includegraphics[width=10cm]{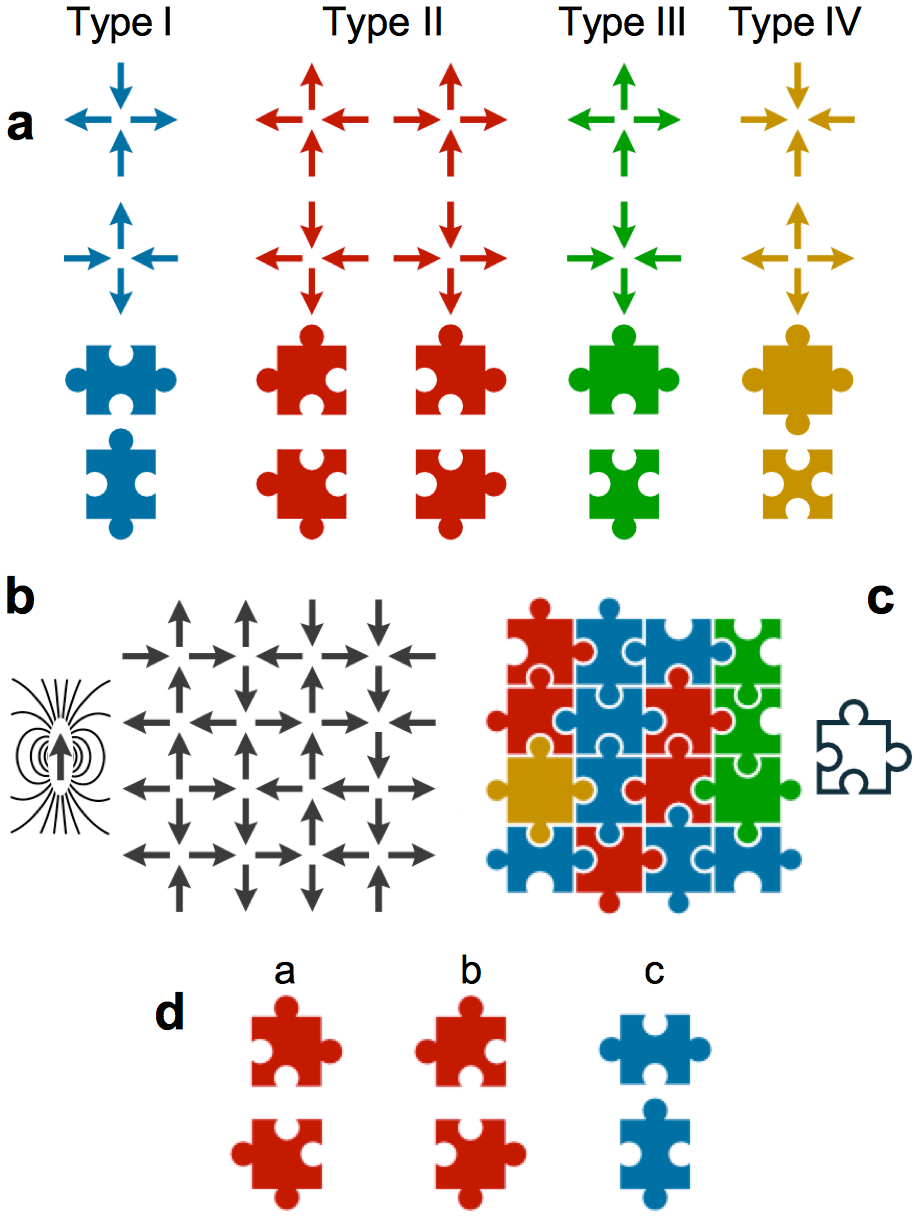}
\caption{\label{vertex} {\bf Vertex types on a square lattice.} (a) All possible vertex configurations. These sixteen different vertices are represented either using the arrows linking neighboring vertices or using a puzzle piece representation to conveniently describe that vertices are non-interacting objects, although they propagate a local constraint. (b) Schematics of a square spin lattice. (c) Puzzle piece representation of a magnetic configuration involving all four vertex types described in (a). (d) Puzzle pieces corresponding to the six possible vertices considered in the six vertex model. Assuming reversal symmetry of the arrows bonding neighboring vertices, three Boltzmann weights $a$, $b$ and $c$ allow the description of the phase diagram associated with the six vertex model.}
\end{figure}

Vertex models are models of statistical mechanics in which a Boltzmann weight is attributed to each vertex of a given lattice.
Although any lattice geometry can be considered in any dimension, we consider in the following the two-dimensional square lattice, which gave rise to celebrated vertex models \cite{Lieb1972, Baxter1982}.
Each of the four links meeting at a vertex is assumed to have two possible states represented by an arrow. 
This arrow can either point inwards or outwards the vertex and can thus be considered as an Ising variable.
The configuration resulting from the four arrows meeting at a vertex $i$ defines its energy $\epsilon_i$.
The total energy $E$ of a given vertex microstate is then the sum of all individual vertex energies: $E=\sum_i \epsilon_i$.
Among the $2^4=16$ possibilities defining a vertex state on a square lattice, six states are made of two spins pointing inwards and two spins pointing outwards the vertex, leading to a local divergence-free condition. 
The remaining ten vertices are states that can be seen as sources or sinks of a magnetic flux, breaking the divergence-free condition. 
These sixteen states are represented in Fig.~\ref{vertex}(a), both in the form of Ising spins and using a puzzle piece representation.

Vertex models differ from their often associated Ising spin models, in which the Boltzmann weight is attributed to the bond connecting two neighboring vertices.
The total energy $E^{'}$ of a given Ising spin microstate is the sum of all pairwise spin configurations: $E^{'}=\frac{1}2 \sum_{(i,j)} J_{ij}~\sigma_i \sigma_j$, where $J_{ij}$ is the coupling strength between the Ising spins $\sigma_i$ and $\sigma_j$ residing on the sites $i$ and $j$, respectively.
One should keep in mind that vertices in vertex models are non-interacting objects, in contrast with spins in spin models [see Fig.~\ref{vertex}(b,c)].
Still, vertex models are constrained models as two neighboring vertices share an Ising variable.
This is illustrated using the puzzle piece representation: a vertex does not directly interact with its four neighbors but imposes a local constraint on each of them. 
So far, artificial spin systems, built from an assembly of magnetic elements interacting through magnetostatics, have been used to capture experimentally the many-body physics associated with frustrated and unfrustrated spin models, but not with the physics of vertex models.

In the following, we consider the different variants of the six vertex model, which restricts the possible vertex states to type I and type II vertices (blue and red puzzle pieces, respectively).
In its general formulation, six different Boltzmann weights are assigned to the six possible vertices \cite{Lieb1972, Baxter1982}.
Assuming reversal symmetry of the arrows bridging neighboring vertices, three Boltzmann weights only allow the description of the system phase diagram \cite{Slater1941, Rys1963, Sutherland1967, Lieb1967a, Lieb1967b, Lieb1967c, Lieb1967d}, which presents four different regions: two ferroelectric phases ($a > b + c$ or $b > a + c$), one antiferroelectric phase ($c > a + b$) and one disordered phase [$a,b,c < \frac{1}2 (a + b + c)$] [see Fig.~\ref{vertex}(d) for the Boltzmann weights $a$, $b$ and $c$].
We will see below how these different phases can be potentially reached by tuning the vertex energy using micromagnetism as a knob.

\subsection{Storing end tuning the energy at the vertex sites of an artificial system}

The central ideal of our work is to consider a square lattice of nanomagnets that we physically connect at the vertex sites. 
Magnetization distribution will then arrange in the form of a domain wall, where the four nanomagnets meet. 
All the energy is then stored at the vertex.
As we will see below, the vertex energy and the energy hierarchy between vertex types can be changed by adding a hole at the vertex site, i.e., by removing some magnetic materials in the array. 
In fact, the internal structure of magnetic domain walls usually results from the competition between the microscopic exchange and the magnetostatic interactions. 
The shape of the nanostructure in which domain walls are confined, together with its geometrical parameters, such as width and thickness, are crucial to determine the internal structure that minimizes the micromagnetic energy \cite{McMichael1997, Nakatani2005, Nguyen2015}. 
The hole that we incorporate within the region where the domain wall seats allows fine tuning of the competition between the microscopic exchange and magnetostatic interactions. 

\begin{figure}[H]
\center
\includegraphics[width=12cm]{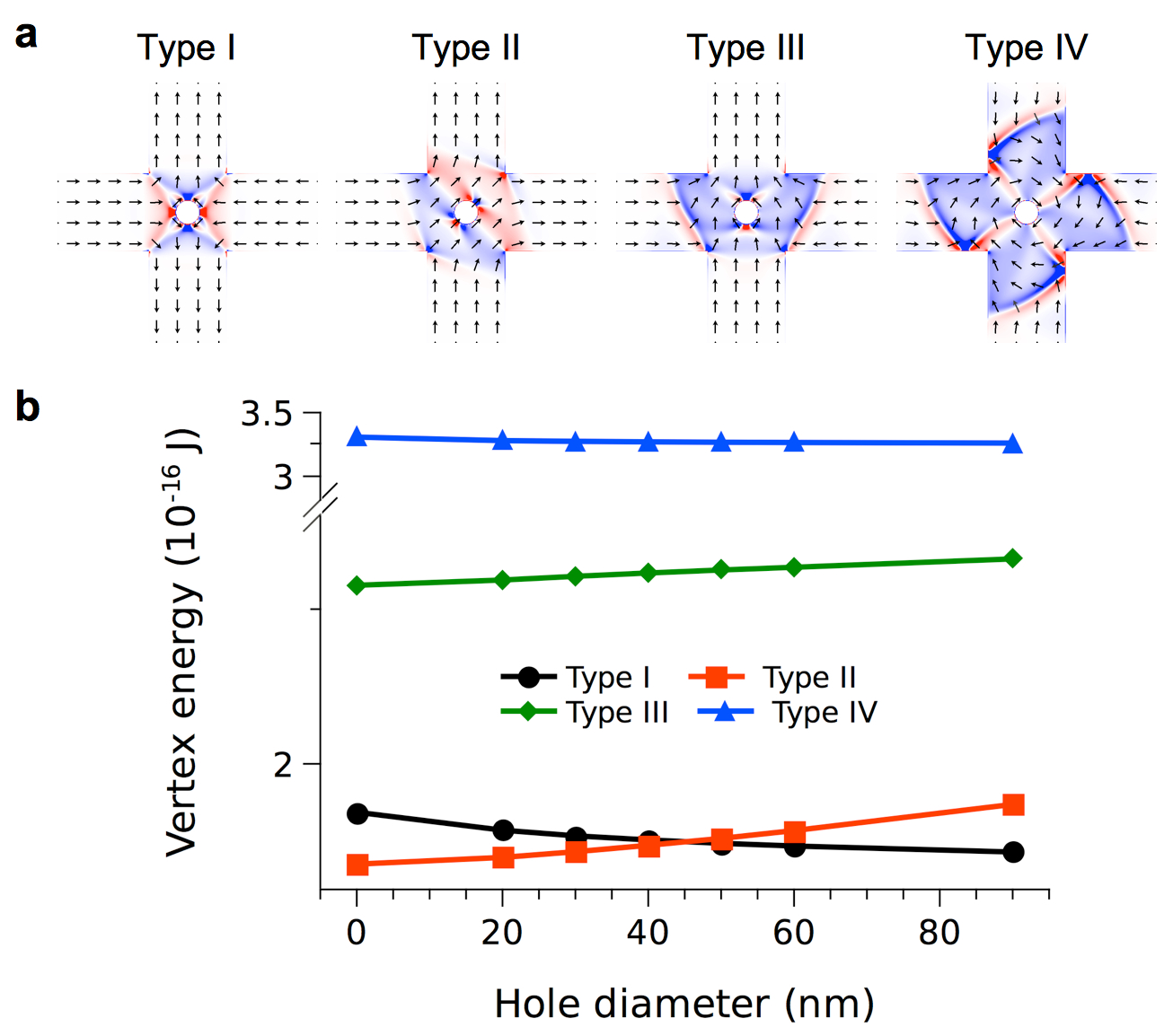}
\caption{\label{micromag} {\bf Using micromagnetic as a knob to modify the vertex energy.} (a) Micromagnetic configurations of type I, type II, type III and type IV vertices for 300 nm-wide, 25 nm-thick permalloy nanomagnets with a hole diameter equals to 90 nm. Black arrows represent the local direction of magnetization, while the blue / red contrast codes for the divergence of the magnetization vector. (b) Micromagnetic energy of the four vertex types when changing the hole diameter for the same geometrical parameters of the nanomagnets.}
\end{figure}

To illustrate this effect, we first use finite difference micromagnetic simulations \cite{OOMMF, mumax} [see Methods].
The micromagnetic configurations for each vertex type are reported in Fig.~\ref{micromag}(a). 
These configurations ressemble domain walls in many ways. 
Type I vertices have the form of a magnetic antivortex, while type II vertices are almost homogeneously magnetized in (11)-like directions. Type III vertices have the form of an enlarged transverse domain wall separating the two horizontal head-to-head nanomagnets. 
Type IV vertices ressemble a vortex domain wall, with two possible chiralities (the vortex core being removed by the presence of the hole). 

In Fig.~\ref{micromag}(b), we report the total energy of the four vertex types as a function of the hole diameter for nanomagnets with typical dimensions of $1500 \times 300 \times 25~nm^3$. 
The main result here is the capability to change the energy hierarchy between type I and type II vertices by adjusting the hole diameter, while keeping type III and type IV vertices much higher in energy.
The fact that the energies of type I and type II vertices vary differently with the hole diameter is explained by the change of the microscopic exchange and magnetostatic energies.
For small hole diameters, the type I vertex costs more energy that the type II vertex because it hosts an antivortex domain wall, a texture known to be highly energetics because of the exchange penalty resulting form the curl of magnetization.
As the hole diameter is increased, the energy of the type II vertex increases because of the additional magnetostatic energy induced by the presence of the hole [as illustrated in Fig.~\ref{micromag}(a)], while at the same time the exchange penalty is reduced for the type I vertex [see Methods].
The crossing point where type I and type II vertices have the same energy then suggests that micromagnetism can serve as a degree of freedom allowing exploration of the different variants of the six vertex model. 
In particular, the disordered phase obtained when the energy is the same for the six vertices should be observed.

\subsection{Experimental results}

To test this idea experimentally, we fabricated a series of square lattices made of permalloy nanomagnets connected at each vertex sites, following the design used for the micromagnetic simulations described above. 
The nanomagnets we consider here are 300 nm-wide and 25 nm-thick. 
The vertex-to-vertex distance is set to 1.8 $\mu$m and the hole diameter is varied between 70 and 210 nm, typically. 
We emphasize that this lattice geometry differs substantially from the one used in previous works on artificial spin systems as it does not consist of an  assembly of interacting magnetic objects. 
Instead, the lattice can be seen as a unique magnetic object.
The square lattices were demagnetized using a field protocol, which efficiently brings arrays of interacting nanomagnets into a low-energy manifold. 
The resulting magnetic configurations are then imaged with a magnetic force microscope (MFM).

\begin{figure}[H]
\center
\includegraphics[width=15cm]{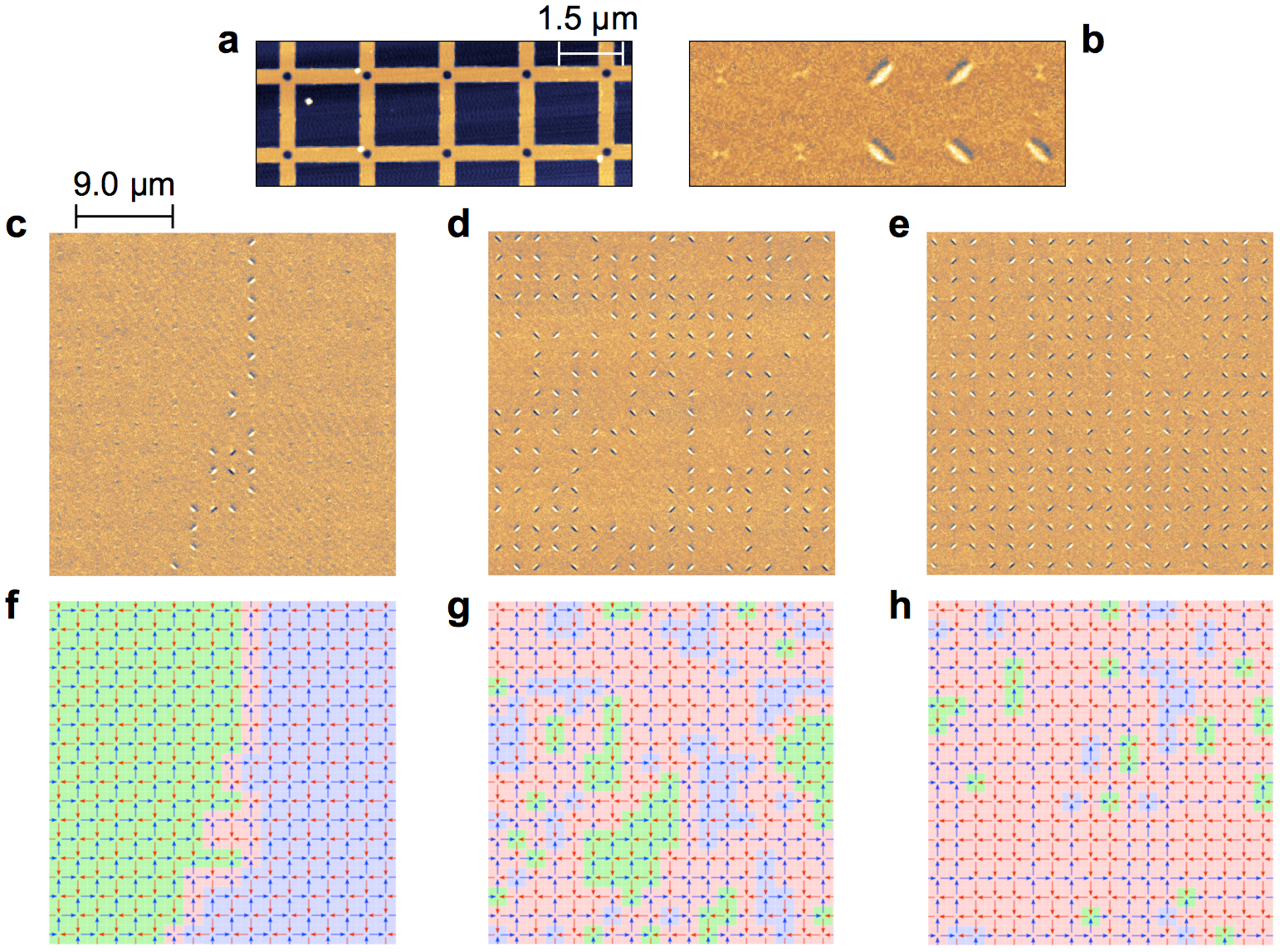}
\caption{\label{MFM} {\bf Experimental results.} (a) Topography image of a connected square lattice with a hole diameter equals to 200 nm (b) Magnetic image corresponding to the topography image reported in (a). Type I and type II vertices can be identified, although type I vertices show a much weaker contrast due to charge screening. Magnetic image of a square lattice for a hole diameter of (c) 210 nm (d) 190 nm and (e) 130 nm. (f-h) Analysis of the spin configuration in (c-e), respectively. Green / blue patches code for regions of the lattice where only type I vertices are found. Red codes for type II vertices.}
\end{figure}

We first observe that the magnetic contrasts associated with type I and type II vertex configurations have different intensity. 
This is illustrated in Figs.~\ref{MFM}(a,b) where this contrast appears much stronger for type II than for type I vertices. 
This simply results from the magnetic charge screening, which is more efficient for a type I vertex as the north (positively charged) and south (negatively charged) poles of the four nanomagnets that meet alternate [see Methods]. 
Consequently, MFM images seem to show at first sight only type II magnetic contrast, but in the regions where the magnetic contrast is barely visible, type I vertices are indeed present.   

We then compare the magnetic configurations of three square lattices having three different hole diameters [see Figures~\ref{MFM}(c-h)]. 
For large hole diameters ($\phi$ = 210 nm), we find the system close to the antiferromagnetic (AFM) ground state configuration built from type I vertices only. 
In Figures~\ref{MFM}(c,f), the lattice shows two large patches of this AFM ground state separated by a domain wall made of type II vertices crossing the entire lattice. 
This result is consistent with predictions from the Rys F model \cite{Rys1963}.
For small hole diameters ($\phi$ = 130 nm), the lattices are found close to the ground state configuration predicted by the KDP model \cite{Lieb1967c}, i.e. a magnetic state made of fully polarized lines crossing the lattice horizontally and vertically. 
Ordering is not perfect and we still observe the presence of type I vertices, diluted within a type II background. 
These diluted type I vertices act as local defects connecting two ferromagnetic lines with opposite directions [see Figure~\ref{MFM}(h)].
Interestingly, for intermediate hole diameters ($\phi$ = 190 nm) the magnetic configuration reveals small patches of both ground states. 
This illustrates that the average size of type I and type II domains can be changed continuously by adjusting the hole diameter, and demonstrates the capability of our approach to change model using micromagnetism as a knob. 
This observation has another important consequence: if properly chosen, the hole diameter could be tuned experimentally to approach the (algebraic) liquid state associated with the disordered phase \cite{Lieb1967a, Lieb1967d}. 

\begin{figure}[H]
\center
\includegraphics[width=7.5cm]{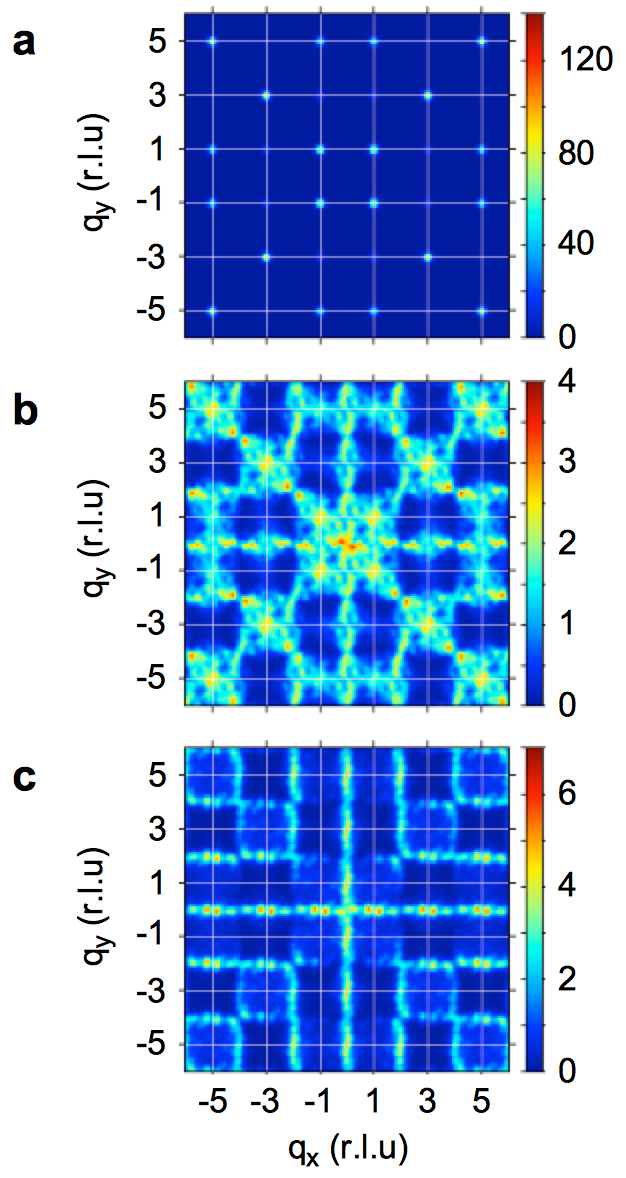}
\caption{\label{structure factor} {\bf Magnetic structure factors deduced from real space magnetic images.} Hole diameter is (a) 210 nm (b) 200 nm (c) 190 nm. The figure illustrates how a fine tuning of the hole diameter modifies the physics of our artificial vertex system. The observed magnetic configurations correspond to a low-energy manifold of (a) the Rys F model (antiferroelectric phase), (b) the ice model (liquid-like disordered phase) and (c) the Slater KDP model (ferroelectric phase).}
\end{figure}

The magnetic structure factor is a convenient tool to characterize the magnetic disorder of a given configuration as it provides a magnetic diffraction pattern representing the spin-spin correlations. 
We thus have computed the magnetic structure factor of three square lattices having different hole diameters, but close to the critical diameter for which the population of type I an type II vertices are comparable to those expected in the disordered phase.
These lattices were demagnetized twice to improve statistics. 
Results are reported in the Figure~\ref{structure factor} for hole diameters $\phi$ = 210 nm [Fig.~\ref{structure factor}(a)], $\phi$ = 200 nm [Fig.~\ref{structure factor}(b)] and $\phi$ = 190 nm [Fig.~\ref{structure factor}(c)].
The magnetic structure factors first confirm our real space observations: tuning the hole diameter allows us to explore several variants of the six vertex model.
For the largest holes, the diffraction pattern is characterized by Bragg peaks at the corners ($M$ direction) of the Brillouin zone [Figure~\ref{structure factor}(a)].
These Bragg peaks are associated with the antiferromagnetic ground state configuration of the Rys F model.
For the smallest hole diameter, the magnetic structure factor reveals a line pattern associated with the ferromagnetically polarized lines observed in real space [Figure~\ref{structure factor}(c)].
This diffraction pattern confirms that our lattices approach in that case the ground state configuration predicted by the KDP model.
Finally, when adjusting properly the hole diameter, the magnetic structure factor reveals a diffuse pattern [Figure~\ref{structure factor}(b)], similar to the one expected within the square ice model \cite{Perrin2016}.
This pattern strongly suggests that our demagnetized square lattices behave as a classical spin liquid.

\subsection{Discussion}

We recall here the three conditions required to have an experimental realization of the disordered phase (at zero temperature) predicted by the six vertex model: 

\noindent 1) The energies of type I and type II vertices must be equal,

\noindent 2) The vertices must be non-interacting objects and the local constraint is only associated with the common bond linking two neighboring vertices, 

\noindent 3) Type III and type IV vertices are not considered (i.e., their occurrence probability is zero). 

\noindent Our system thus offers a promising platform to investigate the disordered phase through a lab-on-chip approach as these three conditions can be very well approximated:

\noindent  1) The hole diameter is an external parameter that can be finely tuned to adjust the energy of type I and type II vertices [see Fig.~\ref{micromag}(b)],

\noindent 2) Vertex-to-vertex interaction can be neglected [see Methods], 

\noindent  3) Monopole-like defects are not observed [see Fig.~\ref{MFM}].

\noindent We emphasize that the second and third conditions are usually not fulfilled when using artificial spin systems, i.e., assemblies of nanomagnets coupled through the magnetostatic interaction.
In particular, although they are prohibited in the square ice model, type III charged defects are very often present in significant amount in artificial spin systems \cite{Perrin2016} as their energy is also reduced in strategies developed to reach the first condition mentioned above. 

However, we note that the critical hole diameter at which we observe a disordered phase differs substantially from the one deduced from the micromagnetic simulations. 
In fact, this critical diameter remains essentially unchanged when varying the size of the nanomagets and is always found of the order of a few tens of nanometers [see Methods].
We attribute these differences to the field demagnetization protocol, which favors ferromagnetic spin-spin correlations within the horizontal and vertical lines of the lattice.
Similar effects have been observed in artificial spin systems \cite{Perrin2016}, but here the effects of the demagnetization protocol might be even stronger since magnetic domain walls mediating magnetization reversal can propagate throughout the lattice, contrary to systems in which the nanomagnets are physically disconnected. 

In addition, careful inspection of the magnetic structure reported in Fig.~\ref{structure factor}(b,c) and the one expected in the disordered phase reveals small differences: the overall background remains slight structured in our lattices and the intensity remains a bit higher than expected.
We believe that the same effects can be evoked here as well, as the demagnetized protocol favors ferromagnetic spin-spin correlations within the lines of the lattices.
Such effects should be strongly limited when using systems that could be thermally activated.
If promising, our design must be optimized if the goal is to explore the properties of the disordered phase, such as the algebraic spin-spin correlations and the pinch points in the associated magnetic structure factor.

\begin{figure}[H]
\center
\includegraphics[width=7.5cm]{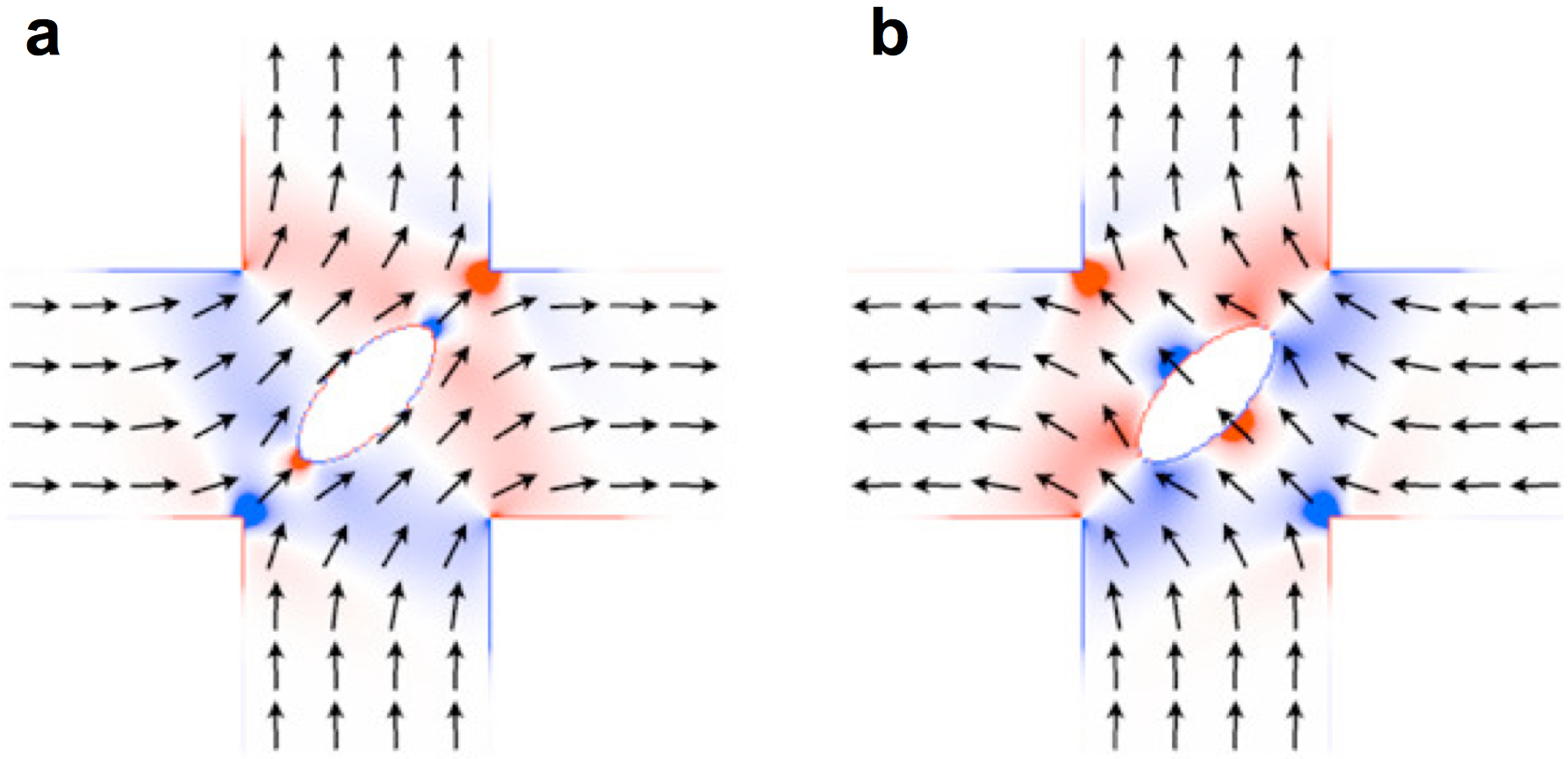}
\caption{\label{ellipses} {\bf Lifting the degeneracy of type II vertices.} Micromagnetic configurations of a type II vertex for 100 nm-wide, 20 nm-thick permalloy nanomagnets with a hole diameter equals to $70 \times 30$ nm$^2$. Two cases are considered depending on how magnetization is aligned with respect to the long axis of the ellipse. Black arrows represent the local direction of magnetization, while the blue / red contrasts code for the divergence of the magnetization vector. The total micromagnetic energies are respectively: (a) $3.33 \times 10^{-17}$ J (b) $3.64 \times 10^{-17}$ J.}
\end{figure}

Regarding the KDP variant of the six vertex model, the design we propose can be easily improved. 
In fact, in the KDP model the degeneracy between the four possible type II vertices is lifted and one of the condition $a > b + c$ or $b > a + c$ should be met.
One elegant way allowing to fulfill this condition is to make the vertex asymmetric by replacing the circular holes by ellipses.
In that case, micromagnetic simulations indicate that the degeneracy between the type II vertices is indeed lifted [see Figure~\ref{ellipses}].
The direction of the ellipse's long axis then determines which of the two conditions is met ($a > b + c$ or $b > a + c$) and can be chosen accordingly.

To conclude, based on the fabrication of lithographically-patterned magnetic lattices, we provide a first attempt to investigate the physics of vertex models experimentally.
The central idea of this work is to replace the spin degree of freedom of artificial frustrated spin systems by a local micromagnetic knob, which can be finely tuned through the design of the vertex geometry.
We believe that our design offers interesting possibilities for studying vertex models through a lab-on-chip approach. 
We hope that our work will stimulate new research in the field and will help connecting experiments with statistical physics.

\subsection{Methods}

\subsubsection{Micromagnetic simulations}
Micromagnetic simulations were performed using the open source OOMMF \cite{OOMMF} and Mumax3 \cite{mumax} softwares. 
Both codes are based on a finite difference approach, i.e., the simulated system is discretized into an orthorhombic mesh.
We computed the micromagnetic energy of the four possible vertex types as a function of the hole diameter located at the vertex site.
In all calculations, the parameters commonly used for permalloy were chosen: spontaneous magnetization $M_S$ = $8 \times 10^5$ A/m (i.e., $\mu_0 M_S = 1.0$ T), the exchange stiffness was set to 10 pJ/m and the magnetocrystalline anisotropy was neglected.
Besides, the damping coefficient was set to 0.5 and magnetic moments at the free extremity of the nanomagnets were fixed to avoid non-uniform magnetization profiles at the edges.
To limit the influence of numerical roughness, the mesh size was set to $2\times 2\times 25$ nm$^3$. 
The nanomagnets have the same dimensions as in the experiments (1500 $\times$ 300 $\times$  25 nm$^3$).

To understand why the hole diameter is changing the energy hierarchy between type I and type II vertices, it is useful to plot the microscopic exchange and demagnetization energies as a function of the hole diameter [see the Supplementary Figure~1]. 
Doing so reveals that these two energies follow the same trend as the hole diameter is increased.
For a type I vertex, they both decrease with the hole diameter, while they both increase for a type II vertex.
However, the change of the exchange energy is about 2.5 larger than the change of the demagnetization energy for a type I vertex, consistently with the fact that the hole is essentially removing the core of the antivortex, where the exchange penalty is high.
For a type II vertex, the situation is reversed and the change of the demagnetization energy is about 7 times larger than the change of the exchange energy.
This is consistent with the fact that the hole is the source in that case of additional magnetic charges.

Considering the dependency of the microscopic exchange and demagnetization energies with the hole diameter, we do not expect any significant change in the critical hole diameter for which type I and type II vertices have the same energy if one changes the typical dimensions of the vertex.
To check this claim, we performed other micromagnetic simulations for nanomagnets with dimensions 500 $\times$ 100 $\times$  20 nm$^3$ and 1000 $\times$ 200 $\times$  20 nm$^3$ [see Supplementary Figure~2].
Indeed, we find that the critical diameter is about 30 and 40 nm, respectively, i.e., close to the 45 nm value reported in Figure~\ref{micromag}(b) for 1500 $\times$ 300 $\times$  25 nm$^3$ nanomagnets.

Using the Mumax3 micromagnetic code, we also computed the expected MFM contrast for a type I and a type II vertex [see the Supplementary Figure~3].
Consistently with what is observed experimentally, a type I vertex exhibits a magnetic contrast sightly weaker than the type II vertex.
We interpret this difference as a consequence of the magnetic charge screening, which is more efficient for a type I vertex, since neighboring nanomagnets have oppositely charged contributions.

Our micromagnetic simulations also reveal that the vertex-to-vertex interaction can be neglected experimentally.
This is illustrated in the Supplementary Figure~4 where the map of the demagnetization field is represented for type I and type II vertices, in the case of conventional arrays of interacting nanomagnets (i.e., artificial spin systems) and of arrays of physically connected nanomagnets (i.e., artificial vertex systems).
In the latter case, the demagnetization field is essentially confined within the array and there is only little magnetic flux outside the vertex.
On the contrary, non-negligible magnetic flux is observed outside the vertex in the case of physically separated nanomagnets.

\subsubsection{Experimental details}
Magnetic images were obtained using a NT-MDT magnetic force microscope and homemade magnetic tips (a 50 nm-thick CoCr alloy is coating the tip of the cantilever). 
Prior to their imaging, the samples were demagnetized using an in-plane magnetic field, oscillating at a 250 mHz frequency, with an amplitude decaying from about 100 mT to zero in several days.
During the demagnetization protocol, the sample is put in rotation at about 20 Hz.

For the magnetic structure factor analysis, several lines of vertices have been removed from the MFM images to discard a potential influence of the boundary conditions. To improve the statistics, the arrays were demagnetized two times and the resulting configurations were averaged.

\subsection{Acknowledgements}

This work was supported by the Agence Nationale de la Recherche through projects no. ANR12-BS04-009 'Frustrated'. The authors acknowledge support from the Nanofab team at the Institut NEEL.

\subsection{Author contributions}
B.C. and N.R. conceived the project. Y. P. was in charge of the sample fabrication. Magnetic imaging measurements were performed by V. S. under the supervision of S. L. D. Data analysis was made by V. S. and N.R. using numerical tools developed by Y. P. All authors contributed to the preparation of the manuscript.

\subsection{Additional information}
Supplementary Information accompanies this paper.

Competing financial interests: The authors declare no competing financial interests.

\newpage

\begin{thebibliography}{2}

\bibitem{Atomes_froids}
Bloch, I., Dalibard, J., \& Zwerger, W. 
Many-body physics with ultracold gases. 
{\it Rev. Mod. Phys.}  {\bf 80}, 885 (2008).

\bibitem{Schiffer_review}
Nisoli, C., Moessner, R. \& Schiffer, P. 
Artificial spin ice: Designing and imaging magnetic frustration.
{\it Rev. Mod. Phys.} {\bf 85}, 1473 (2013).

\bibitem{Tanaka2006}
Tanaka, M., Saitoh, E., Miyajima, H., Yamaoka, T. \& Iye, Y.
Magnetic interactions in a ferromagnetic honeycomb nanoscale network.
{\it Phys. Rev. B} {\bf 73}, 052411 (2006).

\bibitem{Wang2006}
Wang, R. F., Nisoli, C., Freitas, R. S., Li, J., McConville, W., Cooley, B. J., Lund, M. S., Samarth, N., Leighton, C., Crespi, V. H. \& Schiffer, P.
Artificial spin ice in a geometrically frustrated lattice of nanoscale ferromagnetic islands.
{\it Nature} {\bf 439}, 303 (2006).

\bibitem{Pannetier1993}
Runge, K. \& Pannetier, B.
First decoration of superconducting networks.
{\it EuroPhys. Lett.} {\bf 24} 737-742 (1993).

\bibitem{Pannetier2002}
Serret, E., Butaud, P. \& Pannetier, B.
Vortex correlations in a fully frustrated two-dimensional superconducting network.
{\it EuroPhys. Lett.} {\bf 59} 225-231 (2002).

\bibitem{Davidovic1996}
Davidovi\'{c}, D., Kumar, S., Reich, D. H., Siegel, J., Field, S. B., Tiberio, R. C., Hey, R. \& Ploog, K.
Correlations and disorder in arrays of magnetically coupled superconducting rings.
{\it Phys. Rev. Lett.} {\bf 76}, 815 (1996).

\bibitem{Davidovic1997}
Davidovi\'{c}, D., Kumar, S., Reich, D. H., Siegel, J., Field, S. B., Tiberio, R. C., Hey, R. \& Ploog, K.
Magnetic correlations, geometrical frustration, and tunable disorder in arrays of superconducting rings.
{\it Phys. Rev. B} {\bf 55}, 214521 (1997).

\bibitem{Hilgenkamp2005}
Kirtley, J. R., Tsuei, C. C., Ariando, Smilde, H. J. H., \& Hilgenkamp, H.
Antiferromagnetic ordering in arrays of superconducting  $\Pi$-rings.
{\it Phys. Rev. B} {\bf 72}, 6518-6540 (2005).

\bibitem{Libal2006}
Libal, A., Reichhardt, C. \& Reichhardt, C. J. O.
Realizing colloidal artificial ice on arrays of optical traps.
{\it Phys. Rev. Lett.} {\bf 97}, 228302 (2006).

\bibitem{Han2008}
Han, Y., Shokef, Y., Alsayed, A. M., Yunker, P., Lubensky, T. C., \& Yodh, A. G.
Geometric frustration in buckled colloidal monolayers.
{\it Nature} {\bf 456}, 898-903 (2008).

\bibitem{Qi2008}
Y. Qi, T. Brintlinger, and J. Cumings, 
Direct observation of the ice rule in an artificial kagome spin ice.
{\it Phys. Rev. B }{\bf 77}, 094418 (2008).

\bibitem{Morgan2011}
Morgan, J. P., Stein, A., Langridge, S. \& Marrows, C. H.
 Thermal ground-state ordering and elementary excitations in artificial magnetic square ice.
{\it Nature Phys.} {\bf 7}, 75 (2011).

\bibitem{Farhan2013}
Farhan, A., Derlet, P. M., Kleibert,  A., Balan,  A., Chopdekar, R. V., Wyss, M., Perron, J., Scholl, A., Nolting, F. \& Heyderman, L. J.
Direct observation of thermal relaxation in artificial spin ice.
{\it Phys. Rev. Lett.} {\bf 111}, 057204 (2013).

\bibitem{Porro2013}
J. M. Porro, A. Bedoya-Pinto, A. Berger, and P. Vavassori, 
Exploring thermally induced states in square artificial spin-ice arrays.
{\it New J. Phys.} {\bf 15}, 055012 (2013).

\bibitem{Cumings2014}
Cumings, J., Heyderman, L. J., Marrows, C. H., \& Stamps, R. L.
Focus on artificial frustrated systems.
{\it New J. Phys.} {\bf 16}, 075016 (2014).

\bibitem{Kapaklis2014}
V. Kapaklis, U. B. Arnalds, A. Farhan, R. V. Chopdekar, A. Balan, A. Scholl, L. J. Heyderman, and B. Hj\"orvarsson, 
Thermal fluctuations in artificial spin ice.
{\it Nature Nanotech.} {\bf 9}, 514 (2014).

\bibitem{Tierno2016}
Ortiz-Ambriz, A \& Tierno, P.
Engineering of frustration in colloidal artificial ices realized on microfeatured grooved lattices.
{\it Nat. Commun.} {\bf 7}, 10575 (2016).

\bibitem{Perrin2016}
Perrin, Y., Canals, B. \& Rougemaille, N.
Extensive degeneracy, Coulomb phase and magnetic monopoles in artificial square ice.
{\it Nature} {\bf 540}, 410-413 (2016).

\bibitem{Nguyen2017}
Nguyen, V.-D., Perrin, Y., Le Denmat, S., Canals, B. \& Rougemaille, N.
Competing interactions in artificial spin chains.
{\it Phys. Rev. B} {\bf 96}, 014402 (2017).

\bibitem{Rougemaille2011}
Rougemaille, N., Montaigne, F., Canals, B., Duluard, A., Lacour, D., Hehn, M., Belkhou, R., Fruchart, O., El Moussaoui, S., Bendounan, A. \& Maccherozzi, F.
Artificial kagome arrays of nanomagnets: A frozen dipolar spin ice.
{\it Phys. Rev. Lett.} {\bf 106}, 057209 (2011).

\bibitem{Zhang2013}
Zhang, S., Gilbert, I., Nisoli, C., Chern, G.-W., Erickson, M. J., OBrien, L., Leighton, C., Lammert, P. E., Crespi, V. H.  \& Schiffer, P.
Crystallites of magnetic charges in artificial spin ice.
{\it Nature} {\bf 500}, 553 (2013).

\bibitem{Chioar2014-1}
Chioar, I. A., Canals, B., Lacour, D., Hehn, M., Santos Burgos, B., Mente\c{s}, T. O., Locatelli, A., Montaigne, F. \& Rougemaille, N. 
Kinetic pathways to the magnetic charge crystal in artificial dipolar spin ice.
{\it Phys. Rev. B} {\bf 90}, 220407(R) (2014).

\bibitem{Montaigne2014}
Montaigne, F., Lacour, D., Chioar, I. A., Rougemaille, N., Louis, D., Murtry, S. M., Riahi, H., Santos Burgos, B, Mente\c{s}, T. O., Locatelli, A., Canals, B. \& Hehn, M. 
Size distribution of magnetic charge domains in thermally activated but out-of-equilibrium artificial spin ice.
{\it Sci. Rep.} {\bf 4}, 5702 (2014).

\bibitem{Drisko2015}
Drisko, J., Daunheimer, S. \& Cumings, J. 
$FePd_3$ as a material for studying thermally active artificial spin ice systems.
{\it Phys. Rev. B} {\bf 91}, 224406 (2015).

\bibitem{Anghinolfi2015}
Anghinolfi, L., Luetkens, H., Perron, J., Flokstra, M. G., Sendetskyi, O., Suter, A., Prokscha, T., Derlet, P. M., Lee, S. L., \& Heyderman, L. J.
Thermodynamic phase transitions in a frustrated magnetic metamaterial.
{\it Nature Commun.} {\bf 6}, 8278 (2015).

\bibitem{Canals2016}
Canals, B., Chioar, I. A., Nguyen, V.-D., Hehn, M., Lacour, D., Montaigne, F., Locatelli, A., Mente\c{s}, T. O., Santos Burgos, B., \& Rougemaille, N. 
Fragmentation of magnetism in artificial kagome dipolar spin ice.
{\it Nature Commun.} {\bf 7}, 11446 (2016).

\bibitem{Bhat2013}
Bhat, V. S., Sklenar, J., Farmer, B., Woods, J., Hastings, J. T., Lee, S. J., Ketterson, J. B., \& De Long, L. E.
Controlled magnetic reversal in permalloy films patterned into artificial quasicrystals.
{\it Phys. Rev. Lett.} {\bf 111}, 077201 (2013).

\bibitem{Phatak2016}
Brajuskovic, V., Barrows, F., Phatak, C. \& Petford-Long, A. K.
Real-space observation of magnetic excitations and avalanche behavior in artificial quasicrystal lattices.
{\it Sci. Rep.} {\bf 6}, 34384 (2016).

\bibitem{Gilbert2014}
Gilbert, I., Chern, G.-W., Zhang, S., O'Brien, L., Fore, B., Nisoli, C. \& Schiffer, P. 
Emergent ice rule and magnetic charge screening from vertex frustration in artificial spin ice.
{\it Nature Phys.} {\bf 10}, 670 (2014).

\bibitem{Gilbert2015}
Gilbert, I., Lao, Y., Carrasquillo, I., O'Brien, L., Watts, J. D., Manno, M., Leighton, C., Scholl, A., Nisoli, C. \& Schiffer, P. 
Emergent reduced dimensionality by vertex frustration in artificial spin ice.
{\it Nature Phys.} {\bf 12}, 162 (2015).

\bibitem{Zhang2012}
Zhang, S., Li, J., Gilbert, I., Bartell, J., Erickson, M. J., Pan, Y., Lammert, P. E., Nisoli, C., Kohli, K. K., Misra, R., Crespi, V. H., Samarth, N., Leighton, C. \& Schiffer, P. 
Perpendicular magnetization and generic realization of the Ising model in artificial spin ice.
{\it Phys. Rev. Lett.} {\bf 109}, 087201 (2012).

\bibitem{Chioar2014-2}
Chioar, I. A., Rougemaille, N., Grimm, A., Fruchart, O., Wagner, E., Mente\c{s}, T. O., Hehn, M., Lacour, D., Montaigne, F. \& Canals, B.
Nonuniversality of artificial frustrated spin systems.
{\it Phys. Rev. B} {\bf 90}, 064411 (2014).

\bibitem{Chioar2016}
Chioar, I. A., Rougemaille, N. \& Canals, B. 
Ground-state candidate for the classical dipolar kagome Ising antiferromagnet.
{\it Phys. Rev. B} {\bf 93}, 214410 (2016).

\bibitem{Zeissler2013}
Zeissler, K., Walton, S. K., Ladak, S., Read, D. E., Tyliszczak, T., Cohen, L. F. \& Branford, W. R.
The non-random walk of chiral magnetic charge carriers in artificial spin ice
{\it Sci Rep.} {\bf 3}, 1252 (2013).

\bibitem{Rougemaille2013}
Rougemaille, N., Montaigne, F., Canals, B., Hehn, M., Riahi, H., Lacour, D. \& Toussaint, J.-C.
Chiral nature of magnetic monopoles in artificial spin ice
{\it New J. Phys.} {\bf 15}, 035026 (2013).

\bibitem{Lieb1972}
Lieb, E. H. and Wu, F. Y.
Two dimensional ferroelectric models, In: Phase transitions and critical phenomena, ed. by C. Domb and M. S. Green. 
Academic Press, New York, pp. 331-490, 1972.

\bibitem{Baxter1982}
Baxter, R. J.
Exactly solved models in statistical mechanics.
Academic Press, San Diego, 1982.

\bibitem{Slater1941}
Slater, J. C.
Theory of the transition in $KH_2PO_4$.
{\it J. Chem. Phys.} {\bf 9}, 16-33 (1941).

\bibitem{Rys1963}
Rys, F. 
\"{U}ber ein zweidimensionales klassisches konfigurationsmodell.
{\it Helvetica Physica Acta} {\bf 36}, 537 (1963).

\bibitem{Sutherland1967}
Sutherland, B.
Exact solution of a two-dimensional model for hydrogen-bonded crystals.
{\it Phys. Rev. Lett.} {\bf 19}, 103 (1967).

\bibitem{Lieb1967a}
Lieb, E. H. 
Exact solution of the problem of the entropy of two-dimensional ice.
{\it Phys. Rev. Lett.} {\bf 18}, 692-694 (1967).

\bibitem{Lieb1967b}
Lieb, E. H.
Exact solution of the F model of an antiferroelectric.
{\it Phys. Rev. Lett.} {\bf 18}, 1046 (1967).

\bibitem{Lieb1967c}
Lieb, E. H.
Exact solution of the two-dimensional slater KDP model of a ferroelectric.
{\it Phys. Rev. Lett.} {\bf 19}, 108-110 (1967).

\bibitem{Lieb1967d}
Lieb, E. H.
The residual entropy of square ice.
{\it Phys. Rev.} {\bf 162}, 162-172 (1967).

\bibitem{McMichael1997}
McMichael, R. \& Donahue, M.
Head to head domain wall structures in thin magnetic strips.
{\it IEEE Trans. Magn.} {\bf 33}, 4167 (1997).

\bibitem{Nakatani2005}
Nakatani, Y., Thiaville, A. \& Miltat, J. 
Head-to-head domain walls in soft nano-strips: a refined phase diagram.
{\it J. Magn. Magn. Mater.} {\bf 290}, 750 (2005).

\bibitem{Nguyen2015}
Nguyen, V. D., Fruchart, O., Pizzini, S., Vogel, J., Toussaint, J.-C., \& Rougemaille, N.
Third type of domain wall in soft magnetic nanostrips.
{\it Sci. Rep.} {\bf 5}, 12417 (2015).

\bibitem{OOMMF}
Donahue, M. \& Porter, D. Interagency Report NISTIR 6376, National Institute of Standards and Technology, Gaithersburg, MD, 1999.

\bibitem{mumax}
Vansteenkiste, A., Leliaert, J., Dvornik, M., Helsen, M., \& Garcia-Sanchez, F.
The design and verification of mumax3.
{\it AIP Advances} {\bf 4}, 107133 (2014).

\pagebreak
\end{thebibliography}
\end{document}